\documentclass[preprint,12pt]{elsarticle}

\usepackage{graphicx}

\usepackage{amssymb}

\begin{document}

\begin{frontmatter}



\title{Orbital-dependent effects of electron correlations in microscopic models for iron-based superconductors}


\author{Rong Yu}
\address{Department of Physics and Astronomy, Rice University,
Houston, Texas 77005}

\author{Jian-Xin Zhu}
\address{Theoretical Division and Center for Integrated Nanotechnologies,
Los Alamos National Laboratory, Los Alamos, NM 87545}

\author{Qimiao Si\corref{*}}
\cortext[*]{Email: qmsi@rice.edu}
\address{Department of Physics and Astronomy, Rice University,
Houston, Texas 77005}

\begin{abstract}
The bad metal behavior in the normal state of the iron-based superconductors suggests an intimate connection
between the superconductivity and a proximity to a Mott transition.
At the same time, there is strong evidence for the multi-orbital nature of the electronic
excitations. It is then important to understand the orbital-dependent effects of electron correlations.
In this paper we review the recent theoretical progresses on the metal-to-insulator transition
in multiorbital models for the iron-based superconductors. These include studies of models that contain 
at least the 3$d$ $xy$ and 3$d$ $xz/yz$ models, using a
slave-spin technique. For commensurate filling corresponding to that of the parent iron pnictides
and chalcogenideds,
a Mott transition generally exists in all these models. Near the Mott transition, a strongly correlated metal 
exhibiting bad metal features and strong orbital selectivity
is stabilized due to the interplay of Hund's coupling and
orbital-degeneracy breaking.
Particularly for the alkaline iron selenides, the ordered vacancies effectively reduce the kinetic energy,
thereby pushing the system further into the Mott-insulating regime;
in the metallic state,
there exists an orbital-selective Mott phase in which the iron 3$d$ $xy$ orbital is Mott localized while
the other 3$d$ orbitals are still itinerant.
An overall phase diagram for the alkaline iron selenides has been proposed, in which
the orbital-selective Mott phase connects between
the superconducting phase and the Mott-insulating parent state.
\end{abstract}


\end{frontmatter}


\section{Introduction}
\label{Sec:Introduction}

Triggered by the seminal discovery of superconductivity at 26 K in F-doped LaFeAsO~\cite{Hosono}, 
the study on the iron-based high-temperature superconductors has recently become
an important field in condensed matter physics.
The iron-based superconductors consist of two major families of compounds, iron pnictides and iron chalcogenides.
They share a common layered structure, 
with a square lattice of iron ions. The electronic properties of the iron-based superconductors are
closely associated with the iron 3$d$ orbitals.
Typically, the parent compounds have an antiferromagnetically ordered
ground state~\cite{Cruz},
and
superconductivity emerges within a certain range of chemical doping.
The parent systems have the iron valence $+2$, corresponding to 6 electrons occupying
the 3$d$ orbitals.
In the iron pnictides and some iron chalcogenide compounds, the parent
antiferromagnet is metallic. This metallic behavior and the existence of
the hole and electron
Fermi pockets~\cite{Yi09}
raise a central question on the role of the electron correlations.
On the one hand, the antiferromagnetic order and superconductivity may arise from a Fermi surface
nesting mechanism of a weak coupling theory~\cite{Graser09,Ran09,Knolle10}. 
On the other hand, many experiments
imply that the parent pnictides and selenides are bad metals with considerable electron correlations.
This is shown by a large
value of the room-temperature electrical resistivity, which corresponds to a mean-free path of the order of
only $k_F^{-1}$ (Ref.~\cite{AbrahamsSi2011}),
and a sizable reduction of the Drude weight in the optical
conductivity~\cite{Qazilbash,HuOptics,YangOptics,Boris,Degiorgi}.
Indeed, the latter implies that the majority of the electron excitations lie in the incoherent part away from
the Fermi energy.
Furthermore, measurements of the dynamical spin susceptibility
yield an integrated spin spectral weight on the order of 3 $\mu_B^2$ per Fe in the parent 
iron pnictides ~\cite{LiuNatPhys12}. 
This size is consistent
with quasi-local moments generated by the incoherent electronic excitations, but is too large to account for
by the nesting picture of particle-hole excitations from the electrons and holes in the Fermi pockets.
ARPES measurements have identified considerable mass renormalization~\cite{Yi09,Tamai10,Liu13}.
All these suggest that the ratio of the Coulomb interactions $U$ (more precisely, a  combination of the intra-orbital
and inter-orbital Coulomb interactions and the Hund's coupling, see below) to the characteristic bandwidth $W$
is close to $U_c/W$, the threshold for a Mott transition~\cite{SiAbrahams,Haule08}. This
incipient Mott picture
~\cite{SiAbrahams,SiNJP,DaiPNAS}
expands around the Mott transition in terms of the quasiparticle spectral weight
$w$,
which measures the electron correlations in the metallic phase.
It describes a bad metal when
$w
\sim
(U_c-U)/W$ is small. The antiferromagnetic order in the bad metal is associated with the incoherent
electronic
excitations,
which can be modeled by a $J_1$-$J_2$ model of quasi-localized magnetic moments.
Upon doping, the superconductivity arises out of the bad metal on the verge of the Mott localization.
 This can take place when the parent compound is either a bad metal (as for iron pnictides)
 or a Mott insulator (as for alkaline iron selenides).
 A number of studies on the related strong coupling approach have been carried out
 in understanding the electronic properties of iron-based superconductors~\cite{Yin,KSeo,WChen,Moreo,Berg,Lv,Ma,MJHan,Wysocki,Fang:08,Xu:08,Uhrig,Ishida,Lorenzana,Bascones}.

To explore the effect of electron correlations further, it is important to recognize that
the electronic structure of the
iron-based superconductors involves multiple 3$d$ Fe orbitals.
Such multiple orbitals were considered from the 
beginning~\cite{Graser09,SiAbrahams,SiNJP,Kuroki08,Cao08,Haule08,LeeWen08},
and their involvement in the electronic structure near the Fermi energy have been established
by ARPES measurements ~\cite{Yi09}.
The tetragonal structure protects only the degeneracy between the 3$d$ $xz/yz$ orbitals,
but breaks the degeneracy between these orbitals and the remaining three 3$d$ orbitals.
It is therefore instructive to study the orbital-dependent effects of the electron correlations.
The appropriate model will include both the intra- and inter-orbital Coulomb interactions,
as well as the Hund's coupling.

In this paper we report the recent theoretical progress in understanding the bad metal behavior
and the associated
metal-to-insulator transition (MIT) in multiorbital models for iron-based superconductors.
We study both the bad-metal behavior and the metal-to-Mott-insulator transitions
in various multiorbital Hubbard models for the iron pnictides, showing that 
the proximity to the Mott transition becomes strongly orbital-dependent even for
orbitally-independent interactions \cite{Yu_multi1,Yu_multi2}.
Our emphasis is on the interplay among the Hubbard interactions,
Hund's coupling and the orbital non-degeneracy.
For the alkaline iron selenides, we treat the ordered vacancies in terms of a reduction of the bandwidth,
which implies an enhanced $U/W$ ratio \cite{Yu_prl11}.
The corresponding multi-orbital Hubbard models, both in the presence and in the absence of the
ordered vacancies,  are shown to contain an orbital-selective Mott phase (OSMP)~\cite{YuSi12}.
The latter is characterized by localized orbitals (3$d$ $xy$) coexisting with itinerant ones, 
as originally proposed for Ca$_{2-x}$Sr$_{x}$RuO$_{4}$~\cite{Anisimov02}.
The OSMP phase
provides a natural explanation of the strongly orbital- and temperature-dependent
spectral weight observed in the ARPES measurements of the superconducting alkaline iron selenides
~\cite{Yietal12}.
It also provides the understanding of an intermediate phase identified through transport measurements
in the insulating alkaline iron selenides under a large pressure \cite{PGao}.
We note that theoretical studies on related orbital-dependent aspects of electron correlations are being
pursued from a variety of other perspectives and approaches~\cite{Craco11,Yin12,Bascones12,Phillips}.

The rest of the paper is organized as follows: In Sec.~\ref{Sec:Theory} we briefly introduce the multiorbital
Hubbard model for iron-based superconductors, as well as the slave-spin method
\cite{Yu_multi2,FlorensGeorges,deMedici05,deMedici10}
used for studying the MITs in this model. In Sec.~\ref{Sec:MTPnictides} we consider and compare the results
of metal-to-Mott-insulator transition in various multiorbital models for iron pnictides. 
We then proceed to discuss
some recent theoretical studies on alkaline iron selenides in Sec.~\ref{Sec:MTAFS}. 
We address how the ordered iron vacancies enhance the electron correlations and push the system 
into a Mott insulator in the parent alkaline iron selenides,  and discuss the theoretical and experimental 
studies on the OSMP in these systems. 
In Sec.~\ref{Sec:PD_AIS},
we propose an overall phase diagram to understand the combined effects of tuning both the vacancy order 
and carrier doping.
Some general considerations, particularly on what happens 
when the electron occupancy moves away from 6, are given
in Sec.~\ref{Sec:Discussion}. 

\section{Multiorbital Hubbard model and the slave-spin method}
\label{Sec:Theory}

We  consider a multiorbital Hubbard model for iron-based superconductors. The general form of the Hamiltonian reads
\begin{equation}
 \label{Eq:Ham_tot} H=H_0 + H_{\mathrm{int}}.
\end{equation}
Here, $H_0$ contains the tight-binding parameters among the multiple orbitals,
\begin{equation}
 \label{Eq:Ham_0} H_0=\frac{1}{2}\sum_{ij\alpha\beta\sigma} t^{\alpha\beta}_{ij} d^\dagger_{i\alpha\sigma} d_{j\beta\sigma} + \sum_{i\alpha\sigma} (\Delta_\alpha-\mu) d^\dagger_{i\alpha\sigma} d_{i\alpha\sigma},
\end{equation}
where $d^\dagger_{i\alpha\sigma}$ creates an electron in orbital $\alpha$ with spin $\sigma$ at site $i$, $\Delta_\alpha$
is an on-site energy reflecting the crystal field splitting, and $\mu$ is the chemical potential. $H_{\rm{int}}$
contains on-site Hubbard interactions
\begin{eqnarray}
 \label{Eq:Ham_int} H_{\rm{int}} &=& \frac{U}{2} \sum_{i,\alpha,\sigma}n_{i\alpha\sigma}n_{i\alpha\bar{\sigma}}
 +\sum_{i,\alpha<\beta,\sigma} \left\{ U^\prime n_{i\alpha\sigma} n_{i\beta\bar{\sigma}}\right. 
 + (U^\prime-J) n_{i\alpha\sigma} n_{i\beta\sigma}\nonumber\\
&&\left.-J(d^\dagger_{i\alpha\sigma}d_{i\alpha\bar{\sigma}} d^\dagger_{i\beta\bar{\sigma}}d_{i\beta\sigma}
 -d^\dagger_{i\alpha\sigma}d^\dagger_{i\alpha\bar{\sigma}}
 d_{i\beta\sigma}d_{i\beta\bar{\sigma}}) \right\}.
\end{eqnarray}
where $n_{i\alpha\sigma}=d^\dagger_{i\alpha\sigma} d_{i\alpha\sigma}$.
The parameters
$U$, $U^\prime$, and $J$ respectively denote the intraorbital repulsion, the interorbital repulsion, and the Hund's rule exchange coupling. They are taken to satisfy $U^\prime=U-2J$~\cite{Castellani78}.

The electron correlation effects in the multiorbital Hubbard model is studied via the slave-rotor and the slave-spin methods~\cite{Yu_multi2,FlorensGeorges,deMedici05,deMedici10}, which can be viewed as multiorbital versions of
the more standard slave-boson theory~\cite{Kotliar86}.
In the slave-spin representation, the electron creation operator is rewritten
as $d_{i\alpha\sigma}^\dagger=S_{i\alpha\sigma}^+ f_{i\alpha\sigma}^\dagger$. Here, $S_{i\alpha\sigma}^+$
is the ladder operator of a quantum S = 1/2 slave spin that carries the electric charge for each orbital and spin flavor,
and $f_{i\alpha\sigma}^\dagger$ creates an auxiliary fermion (the spinon) that carries the spin of the electron.
The restriction of the Hilbert space to the physical subspace is accomplished by enforcing
a constraint, $S_{i\alpha\sigma}^z=f_{i\alpha\sigma}^\dagger f_{i\alpha\sigma} - 1/2$,
on each site. A mean-field theory can be obtained by first rewriting the multiorbital Hamiltonian in terms
of the slave spins and spinons, then decoupling them. 
Further performing a single-site approximation and taking into account the
translational symmetry of the system in the paramagnetic phase, the resulting mean-field Hamiltonians read:
\begin{eqnarray}
 \label{Eq:Hfmf}  H^{\mathrm{mf}}_f &=&  \sum_{k\alpha\beta}\left[ \epsilon^{\alpha\beta}_{k} 
 \langle O^\dagger_\alpha \rangle \langle O_\beta \rangle + \delta_{\alpha\beta}
 (\Delta_\alpha-\lambda_\alpha+\tilde{\mu}_\alpha-\mu)\right] f^\dagger_{k\alpha} f_{k\beta}, \\
 \label{Eq:HSSmf} H^{\mathrm{mf}}_{S} &=& \sum_{\alpha\beta} \left[\epsilon^{\alpha\beta}
 \left(\langle O^\dagger_\alpha\rangle O_\beta+ \langle O_\beta\rangle O^\dagger_\alpha\right)
 + \delta_{\alpha\beta}\lambda_\alpha S^z_\alpha\right] + H_{\mathrm{int}}(\mathbf{S}).
\end{eqnarray}
Here the spin index $\sigma$ has been dropped for simplicity. $\epsilon^{\alpha\beta}_{k}=\frac{1}{N}\sum_{ij} 
t^{\alpha\beta}_{ij} e^{ik(r_i-r_j)}$, $\epsilon^{\alpha\beta} 
= \sum_{ij\sigma} t^{\alpha\beta}_{ij} \langle f^\dagger_{i\alpha\sigma}
f_{j\beta\sigma}\rangle/2$, $\delta_{\alpha\beta}$ is the Kronecker's
delta function, and $\lambda_\alpha$ is the Lagrange multiplier to handle the constraint. 
In addition, $\tilde{\mu}_\alpha$ is an effective chemical potential which is introduced 
to recover the correct noninteracting limit~\cite{YuSi12}, $H_{\mathrm{int}}(\mathbf{S})$ 
refers to the interaction Hamiltonian in the slave-spin representation~\cite{Yu_multi1},
and
$O^\dagger_\alpha=\langle P^+_\alpha\rangle S^+_\alpha \langle P^-_\alpha\rangle$,
where $P^\pm_\alpha=1/\sqrt{1/2+\delta\pm S^z_\alpha}$, with $\delta$ being
an infinitesimal positive number to regulate $P^\pm_\alpha$. 
The quasiparticle spectral weight $Z_\alpha=|\langle O_\alpha \rangle|^2$.

The formalism of the slave-rotor theory is very similar, but the slave particle that carries the charge 
degree of freedom in this case is a quantum $O(2)$ rotor. 
It is efficient if the interaction has an $SU(2M)$ symmetry for $M$ degenerate orbitals, 
{\it i.e.} when $J=0$ in Eq.~\ref{Eq:Ham_int}. 
On the other hand, the slave-spin method is more convenient to handle systems with finite Hund's rule 
coupling and those exhibiting a strong orbital dependence.

Both the slave-rotor and slave-spin theories can describe the MIT in the multiorbital Hubbard model. 
In particular, the metallic phase corresponds to the state in which the slave particles are Bose condensed ($Z>0$), 
so that charge excitations are gapless along with the spin excitations.
The Mott insulator corresponds to the state in which the slave particles are disordered ($Z=0$);
here the charge excitations are gapped, while the spin excitations remain gapless.

\section{Metal-to-insulator transition in multiorbital models for iron pnictides}
\label{Sec:MTPnictides}

Theoretically, an important question is whether a Mott insulator can generally be stabilized in the multiorbital
models  for the parent iron pnictides.
The answer is not obvious because the parent iron pnictide compounds contain an even number (six)
of electrons per iron ion, and the insulating state could be a band insulator. To fully address this issue, the MIT
has been investigated theoretically using the slave-spin mean-field theory in a various multiorbital models,
from a minimal two-orbital model~\cite{Raghu08} to the more realistic four- and
five-orbital models~\cite{Yu_multi1,Graser09}. The transition is studied in the paramagnetic phase
with a tetragonal lattice symmetry, to highlight the (incipient) localization effects of the Coulomb repulsive interactions.
It is found that a Mott transition generally exists in all these models, and the insulating phase is a Mott insulator.

This result relies on the degeneracy of the $xz$ and $yz$ orbitals. For electron filling corresponding
to the parent compound, they form both the hole-like bands near the $(0,0)$ point and and the electron-like bands
near the $(\pi,0)$ and $(0,\pi)$ points of the Brillouin zone of the one iron unit cell. These bands cross the Fermi energy, respectively giving rise to the hole and electron Fermi pockets. As interaction is turned on, the bandwidths are renormalized. Due to the orbital dependence of $Z_\alpha$, bands with a mixed orbital character is distorted from the noninteracting case. However, since the orbital characters at the bottom of the electron bands (at $(\pi,0)$ and $(0,\pi)$) and the top of the hole bands (at $(0,0)$) are both degenerate $xz$/$yz$, they are renormalized by the same factor. As a result, the bottom of the electron bands are always in lower energy than the top of the hole band, and a finite Fermi surface with $xz$/$yz$ orbital characters survives. Hence a band insulator cannot be stabilized, and the system stays in the metallic state until
$Z_{xz/yz}=0$, where it undergoes a transition to a Mott insulator. The Mott insulator is adiabatically connected
to the atomic limit configuration, where the degenerate $xz$/$yz$ orbitals are half filled in the parent compound.

\begin{figure}[h]
\centering
\includegraphics[totalheight=0.25\textheight
, viewport=50 80 500 360,clip
]{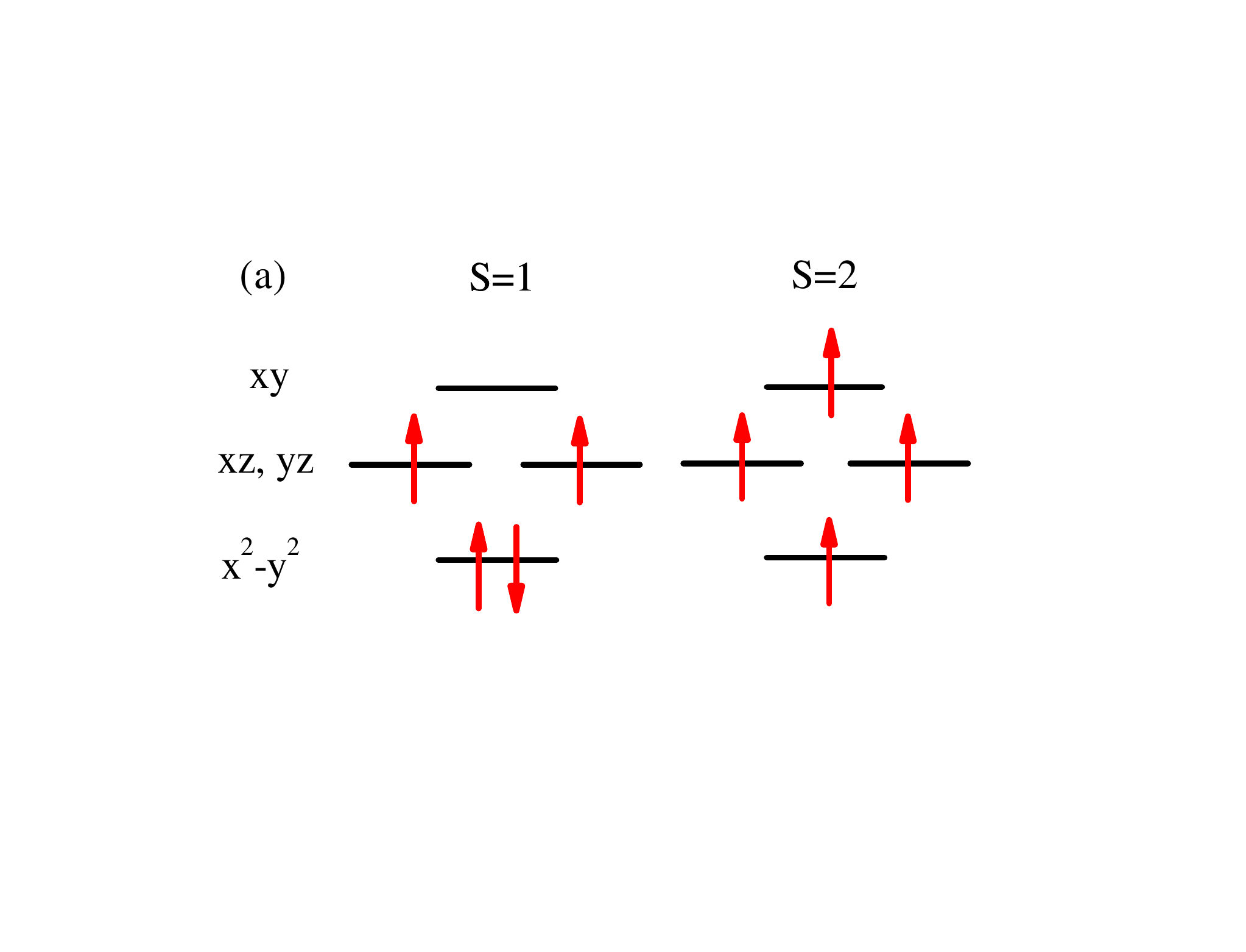}%
\includegraphics[totalheight=0.3\textheight
, viewport=0 0 550 480,clip
]{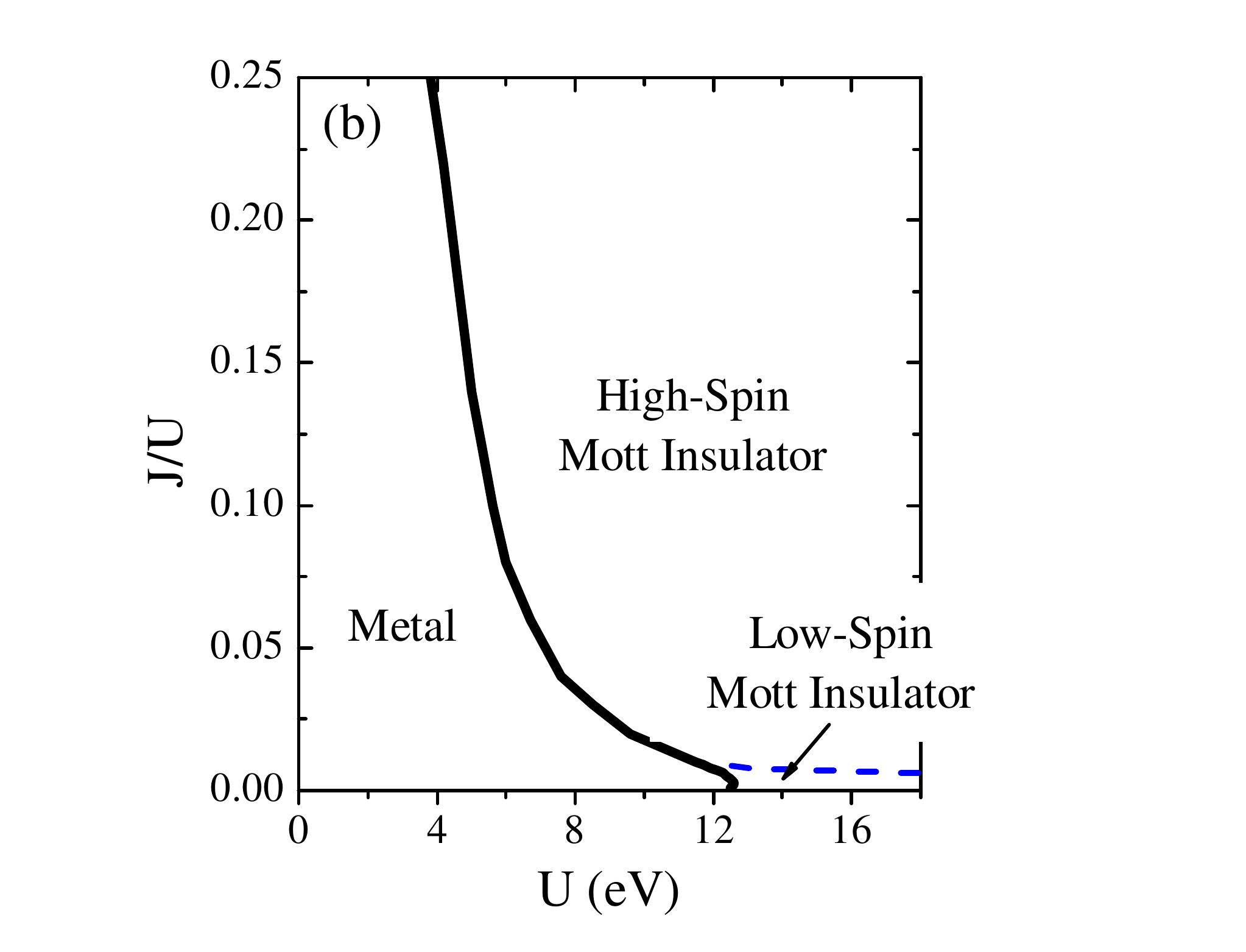}%
\vskip -1pc
\caption{\label{fig2} (a): The $S=1$ low-spin and $S=2$ high-spin configurations at commensurate filling
$N=4$ in the atomic limit of the four-orbital Hubbard model for iron pnictides.
(b): Ground-state phase diagram of the four-orbital model in the slave-spin mean-field theory.
The black solid line corresponds to the metal-to-Mott-insulator transition,
and the blue dashed line refers to a low-spin to high-spin transition between the two Mott insulators.
Adapted from Ref.~\cite{Yu_multi1}.
 }
\end{figure}

\begin{figure}[h]
\centering
\includegraphics[totalheight=0.5\textheight
]{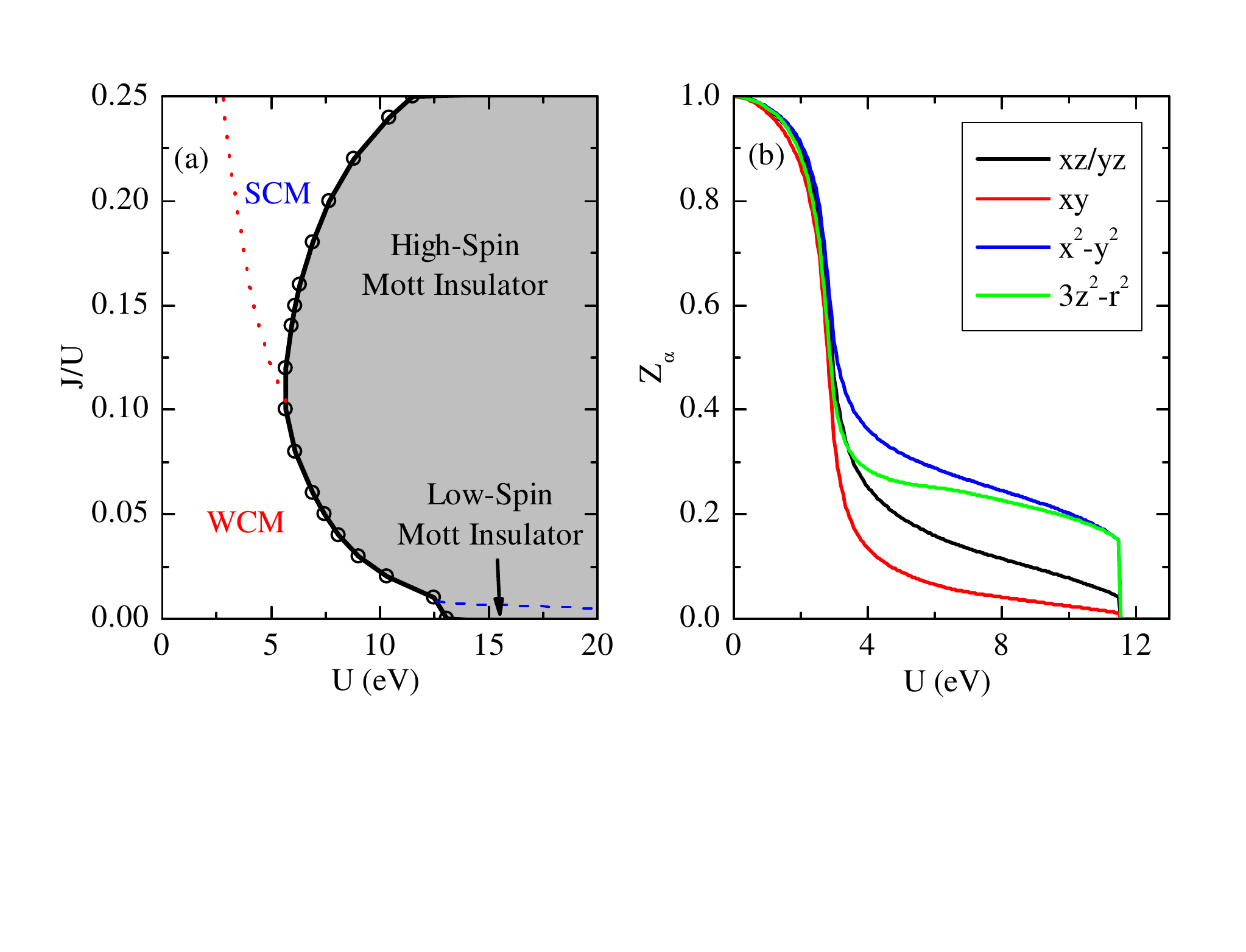}%
\vskip -6pc
\caption{\label{fig3} (a): Ground-state phase diagram of the five-orbital Hubbard model for iron pnictides 
at commensurate filling $N=6$ in the slave-spin mean-field theory. The black solid line with circles corresponds 
to the metal-to-Mott-insulator transition, the blue dashed line refers to a low-spin to high-spin transition 
between the two Mott insulators, and the red dotted line indicates a crossover between the weakly 
correlated metal (WCM) and the strongly correlated metal (SCM). 
(b): The evolution of the orbital resolved quasiparticle spectral weights with $U$ in the five-orbital model 
at $J/U=0.25$. Adapted from Ref.~\cite{Yu_multi2}.
 }
\end{figure}

In multiorbital systems, the Hund's coupling $J$ may affect the nature of both the Mott insulating and metallic states.
This is seen by comparing the results in a four- and a five-orbital system.
There is a finite crystal field splitting $\Delta$ between the highest and lowest orbitals in these models.
Hence, depending on the values of $J$ and $\Delta$, the Mott insulator can be either an $S=1$ low-spin
or an $S=2$ high-spin state, respectively corresponding to the ground-state configurations shown in Fig.~\ref{fig2}(a).
There is a low-spin to high-spin transition between these two Mott states at $J\sim\Delta$ \cite{Yu_multi1}.
As shown in the phase diagrams in Figs.~\ref{fig2}(b) and ~\ref{fig3}(a), for both the four- and five-orbital models,
the low-spin Mott state is stabilized for $J/U\lesssim0.02$. In general, the $S=1$ low-spin Mott state
can be viewed as a mixture of an $S=1$ Mott insulator for the degenerate $xz$ and $yz$ orbitals,
and an orbitally polarized insulator for
  the non-degenerate $xy$ and $x^2-y^2$ orbitals that originates from the interplay of the crystal field splitting
  and the pair hopping interaction~\cite{Yu_multi1}.

For nonzero Hund's couplings, the MIT is generally first-order, and the phase boundary between the metallic
and insulating states is sensitive to the Hund's coupling. For $J/U\lesssim0.1$, in both the four- and
five-orbital models, the critical value $U_c$ of the MIT decreases with increasing $J/U$.
This is because the Hund's coupling suppresses orbital fluctuations by lifting the degeneracy of ground-state
configurations, and effectively reduces the kinetic energy. For $J/U\gtrsim0.1$, while $U_c$ still monotonically
decreases with $J/U$ in the four-orbital model, it increases with $J/U$ in the five-orbital model.
This difference arises from the different behaviors of the Mott gaps in these systems. In the four-orbital model,
which is at half-filling, the Mott gap $G\sim U+3J$. Hence $U_c$ decreases with $J$.
On the other hand, the five-orbital model is away from half-filling. The Mott gap $G\sim U-3J$,
so $U_c$ increases with $J$.

Besides the MIT, there is another crossover at $U^\star$ for $J\gtrsim0.1$ in the five-orbital model.
This crossover line divides the metallic phase into two regimes (Fig.~\ref{fig3}(a)): a weakly correlated metal (WCM)
in which the quasiparticle spectral weights are close to $1$, and a strongly correlated metal (SCM)
in which the quasiparticle spectral weights are significantly suppressed and are strongly orbital dependent.
It is found that an OSMP is energetically in competition with the true ground state~\cite{Yu_multi2}.
The existence of the SCM originates from a combined effect of the Hund's coupling and the crystal field splitting.
As just discussed above, the Hund's coupling lifts the degeneracy of ground-state configurations and reduces
the kinetic energy. In the SCM, the dominant configuration is the high-spin one.
This reduces the correlations among orbitals, making them almost decoupled from each other.
Due to the crystal field splitting,
the filling factor is different in
  each orbital. As a result, the quasiparticle weights become strongly orbital dependent.
  The SCM has all the features of a bad metal in the strong coupling theory. In models for iron pnictides,
  it naturally takes a multiorbital nature.

\section{Mott localization and orbital-selective Mott phase in multiorbital models for alkaline iron selenides}
\label{Sec:MTAFS}

\begin{figure}[t!]
\centering
\includegraphics[totalheight=0.5\textheight
]{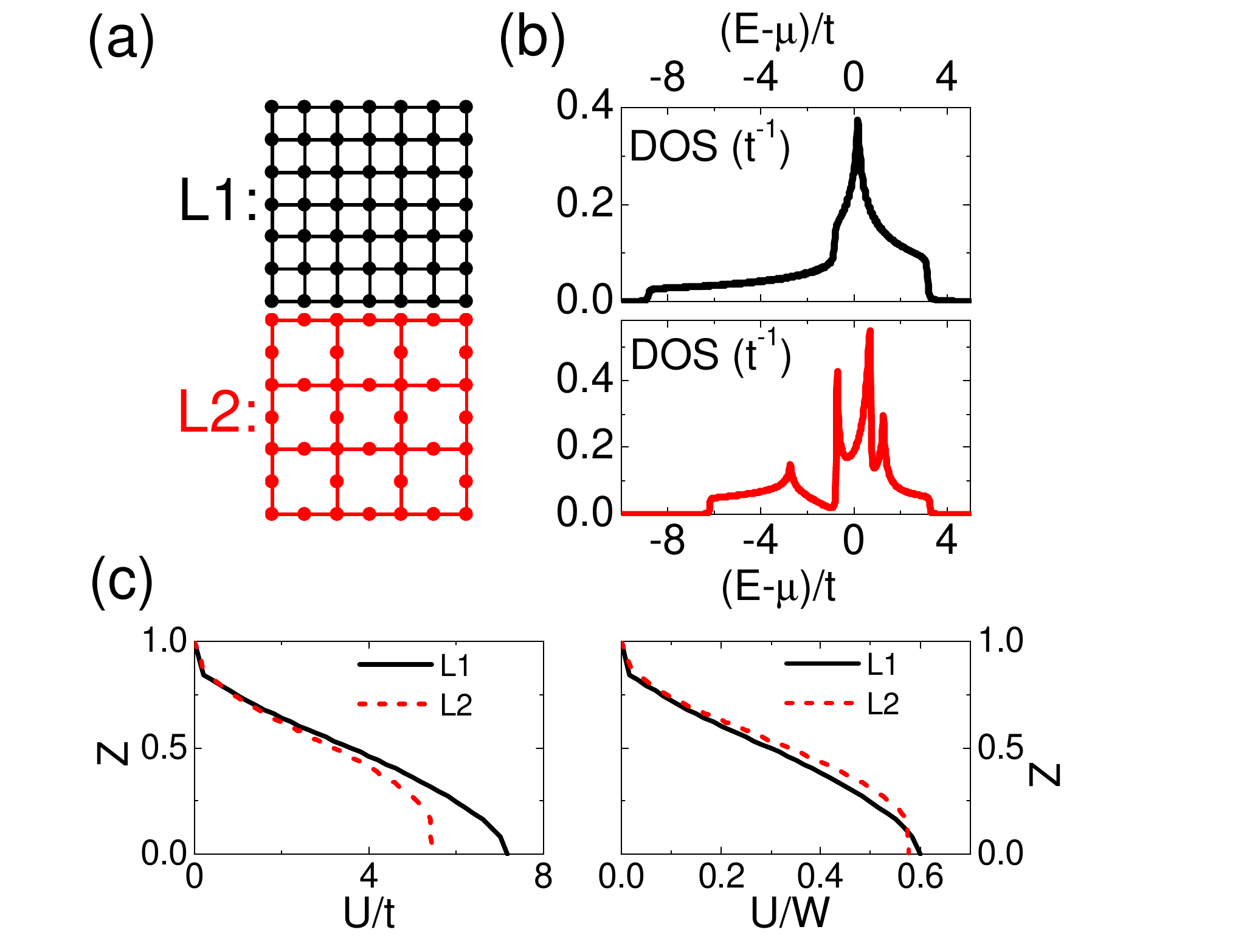}%
\vskip -1pc
\caption{\label{fig4} The Mott localization driven by the iron vacancy order in alkaline iron selenides. (a): The regular (L1) and $1/4$-depleted (L2) square lattices respectively representing the disordered and ordered iron vacancies (The vacancy ordered pattern is chosen for illustrative purpose, such that the model calculation is simplified, and similar results arise for other patterns); (b): Density of states (DOS) of the tight-binding model on the two lattices; (c): Evolution of the quasiparticle spectral weight with $U$, indicating the Mott threshold $U_c$ tracks the bandwidth $W$. $t$ refers to the nearest-neighbor hopping parameter. Adapted from Ref.~\cite{Yu_prl11}.
}
\end{figure}

The recently discovered alkaline iron selenides have some unique features among the iron-based 
superconductors~\cite{Guo,MFang,Krzton-Maziopa,Mizuguchi}. The superconducting compounds have 
only electron Fermi pockets~\cite{Zhang_Feng,Qian_Ding,Mou}, but the superconducting transition temperature 
$T_c$ is comparable to that of iron pnictides~\cite{Guo,Sun12}. Moreover, the superconducting compounds 
are close to a parent insulating state with ordered iron vacancies~\cite{MFang,Wang1101_0789,Bao245} 
and a large antiferromagnetic ordering moments around $3\mu_B$ per iron ion~\cite{Bao245,WangDai11}. 
All these suggest strong electron correlations.

How to connect the Mott insulating state of the parent alkaline iron selenides and the bad-metal nature 
of the parent
iron pnictides? A natural idea is that the former has been driven to the Mott insulating side because 
the ordered
vacancies have reduced the kinetic energy, in a way that is similar to what happens under an 
expansion
of lattice ~\cite{Zhu10}. Towards this end, an effective two-orbital model on two iron lattices 
has been
studied to understand the nature of the iron vacancy ordered parent insulating state~\cite{Yu_prl11}.
As illustrated in Fig.~\ref{fig4}(a), lattice L1 refers to a regular square lattice,
corresponding to the disordered iron vacancies; lattice L2 is a depleted square lattice,
corresponding to the ordered iron vacancies. A remarkable observation is that the ordered 
vacancies
lead to a band narrowing (Fig.~\ref{fig4}(b)). This effectively reduces the kinetic energy and 
pushes the system to the Mott insulator side. Mott localization driven by the band-narrowing effect 
is also found in the
 iron-based material La$_2$O$_2$Fe$_2$O(Se,S)$_2$, in which the band-narrowing is induced
 by the expanded lattice structure~\cite{Zhu10}.

The existence of the Mott-insulating phase in the presence
of vacancy order in K$_{0.8}$Fe$_{1.6}$Se$_2$
(the so-called ``245") alkaline iron selenides suggests that the underlying
correlations are strong in the alkaline iron selenides. At the same time, because vacancy order
represents only about 30\% reduction in the bandwidth $W$ of the 245 alkaline iron selenides
compared to the iron arsenides, the existence of the Mott-insulating phase implies that already
in the iron pnictides, $U/W$ is less than 30\% away from the Mott transition threshold
$U_c/W$; in other words, the correlations in the iron arsenides are also sufficiently strong to place
these materials in proximity to the Mott transition.
 In this way, the importance of the insulating state of the vacancy-ordered alkaline iron selenides
 extends beyond this specific family; it in fact provides evidence for the
incipient Mott picture of the metallic iron arsenides and iron selenides.

\begin{figure}[h]
\centering
\includegraphics[totalheight=0.5\textheight
]{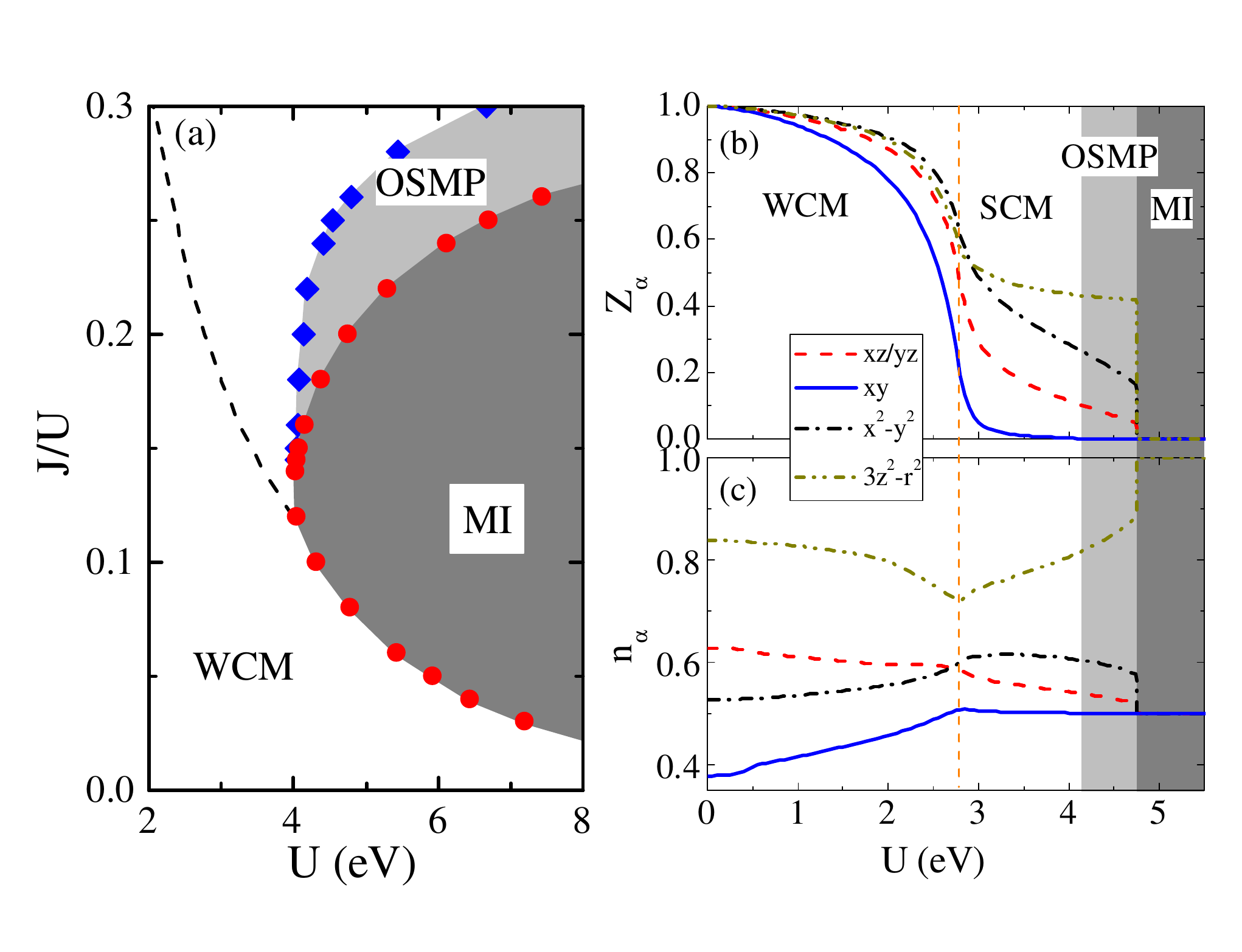}%
\vskip -1pc
\caption{\label{fig5} (a): Ground-state phase diagram of the five-orbital model for alkaline iron selenides 
at commensurate filling $N=6$. The dark and light grey regions correspond to the Mott insulator (MI) 
and orbital-selective Mott phase (OSMP), respectively. The black dashed line refers to a crossover 
between the weakly correlated metal (WCM) and the strongly correlated metal (SCM). 
(b) and (c): The evolution of the orbital resolved quasiparticle spectral weights (in (b)) and electron filling factor 
per spin (in (c)) with $U$ in the five-orbital model at $J/U=0.2$. Adapted from Ref.~\cite{YuSi12}.
 }
\end{figure}

The Mott localization in parent alkaline iron selenides is further confirmed in a more realistic five-orbital model 
by using the slave-spin method~\cite{YuSi12}. The results are illustrated in Fig.~\ref{fig5}.
Besides the Mott insulator, this model also stabilizes an OSMP when the Hund's coupling is beyond a threshold. 
In the OSMP,  the iron 3$d$ $xy$ orbital is Mott localized, while the other 3$d$ orbitals  are still itinerant. 
Several factors favor the OSMP. In the noninteracting limit, the bandwidth projected to the $xy$ orbital is narrower 
than those of other orbitals. Due to the crystal field splitting, the non-degenerate $xy$ orbital lies higher than others, 
making it easier to pin the $xy$ orbital at half-filling. The Hund's coupling suppresses inter-orbital correlations;
together with the crystal field splitting, it effectively decouples the $xy$ orbital from others. 
As a combined effect of all these factors, the $xy$ orbital has a lower interaction threshold for the Mott transition.

\begin{figure}[h]
\centering
\includegraphics[totalheight=0.3\textheight
]{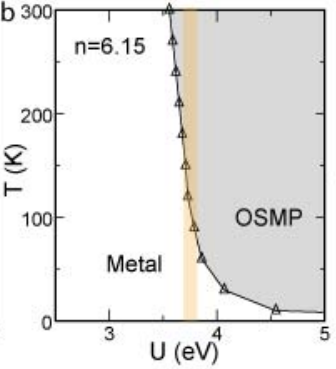}%
\includegraphics[totalheight=0.3\textheight
]{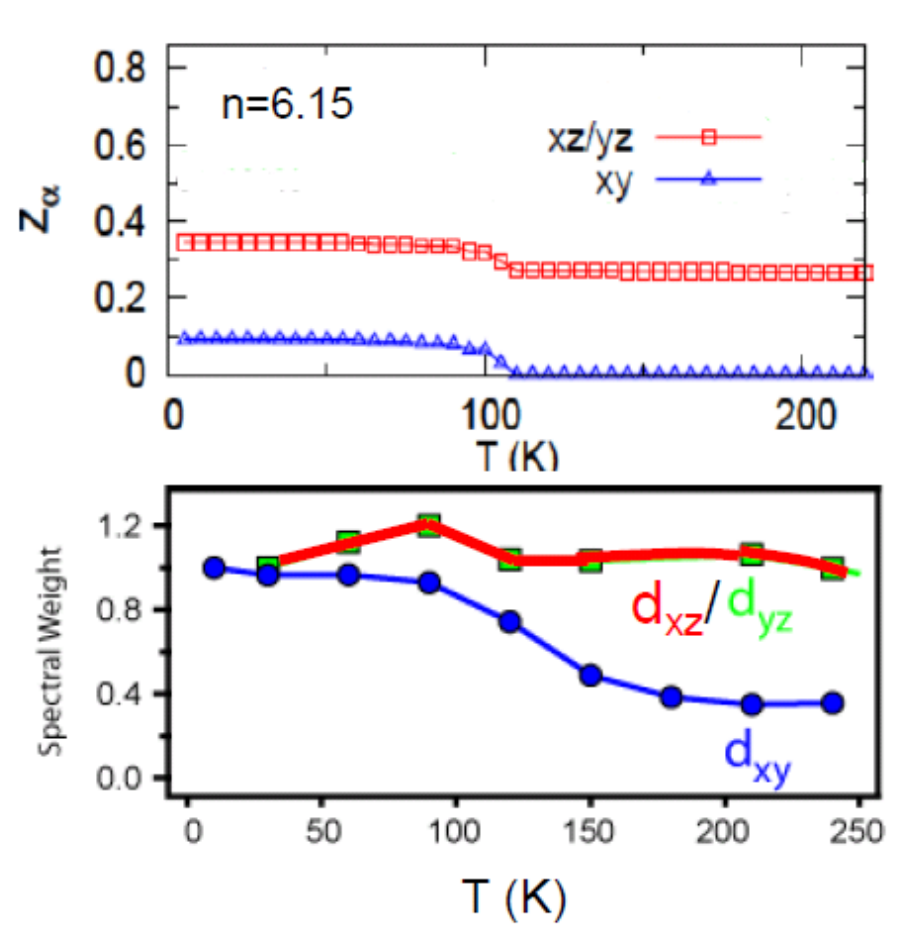}%
\vskip -1pc
\caption{\label{fig6} Orbital-selective Mott transition induced by temperature. (a): The phase diagram 
of the five-orbital model at electron filling $N=6.15$ per iron. (b) and (c): Evolution of quasiparticle spectral weights 
of $xz/yz$ and $xy$ orbitals with temperature calculated from slave-spin theory (in (b)) and observed 
in ARPES measurements (in (c)). Adapted from Ref.~\cite{Yietal12}.
 }
\end{figure}

This OSMP survives finite electron doping. Recently the transition to this OSMP induced by increasing temperature 
has been observed by ARPES measurements~\cite{Yietal12}. The strong suppression of the quasiparticle 
spectral weight in the $xy$ orbital is consistent with the theoretical results, as shown in Fig.~\ref{fig6}. 
Recent high-pressure experiments on the insulating alkaline iron selenides find an intermediate metallic phase 
once the iron vacancy order is fully suppressed by the applied pressure~\cite{PGao}. 
The resistance in this intermediate phase shows an unconventional kink feature, which can also be 
understood by interpreting the intermediate phase as an OSMP. The theoretical analysis together 
with the experimental evidences suggest the OSMP is a necessary intermediate state that 
connects the superconducting alkaline iron selenide with its vacancy ordered Mott insulating parent compound.

\section{An overall phase diagram for alkaline iron selenides}
\label{Sec:PD_AIS}

There is considerable evidence that, in the alkaline iron selenides,
the vacancy-ordered insulating state and
vacancy-free superconducting state
are phase-separated.
If the primary effect of the ordered vacancies on the electronic properties
is to reduced the kinetic energy, one could consider the tuning of the degree
of the ordered vacancies as a means to tune $U/W$. This motivates the consideration
of an overall phase diagram for the alkaline iron selenides, and a physical pathway that
may connect the insulating and superconducting phase.

\begin{figure}[t!]
\centering
\includegraphics[totalheight=0.5\textheight
]{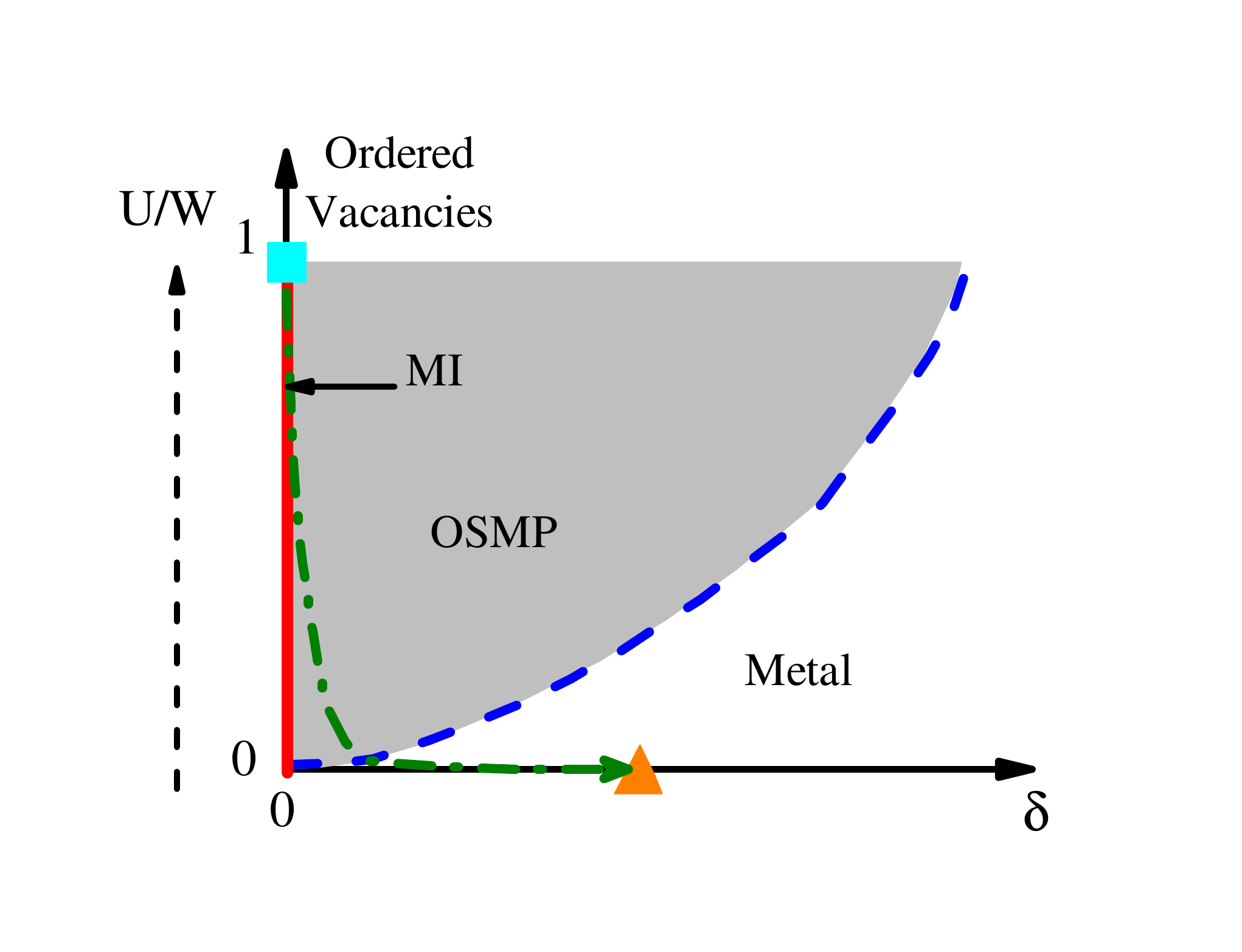}%
\vskip -1pc
\caption{\label{fig8} Sketch of a material-based phase diagram in the plane of carrier doping $\delta$ 
and ordered vacancies. The vacancy order parameter has been scaled to be between 0 and 1. 
The vacancy ordered insulating state is located as the cyan square in this phase diagram. 
The superconducting state is tentatively placed as the orange triangle on the $\delta$ axis. 
The dash-dotted line shows a possible route to connect the two phases. For realistic parameters, 
the Mott transition point is close to the origin, but
could be either above or below it. Adapted from Ref.~\cite{YuSi12}.
 }
\end{figure}

To be more specific here, both the vacancy-ordered 245 insulating phase
and the vacancy-free superconducting phase
come from an underlying multi-orbital Hamiltonian, one in the presence of the
ordered vacancies and the other in the absence of the vacancies.
The existence of the multiple phases in the alkaline iron selenides
suggests the existence of an overall, extended, parameter space
in which the different phases can be connected. The calculations outlined in the previous section have
motivated us to propose~\cite{YuSi12} such an phase diagram, which is shown in Fig.~\ref{fig8}.
This phase diagram  invokes the combined parameter space of vacancy order and carrier
doping. In it, the orbitally-selective Mott phase serves as the link between the parent
insulating phase and the  carrier-doped metallic/superconducting phase.

\begin{figure}[t!]
\centering
\includegraphics[totalheight=0.5\textheight
]{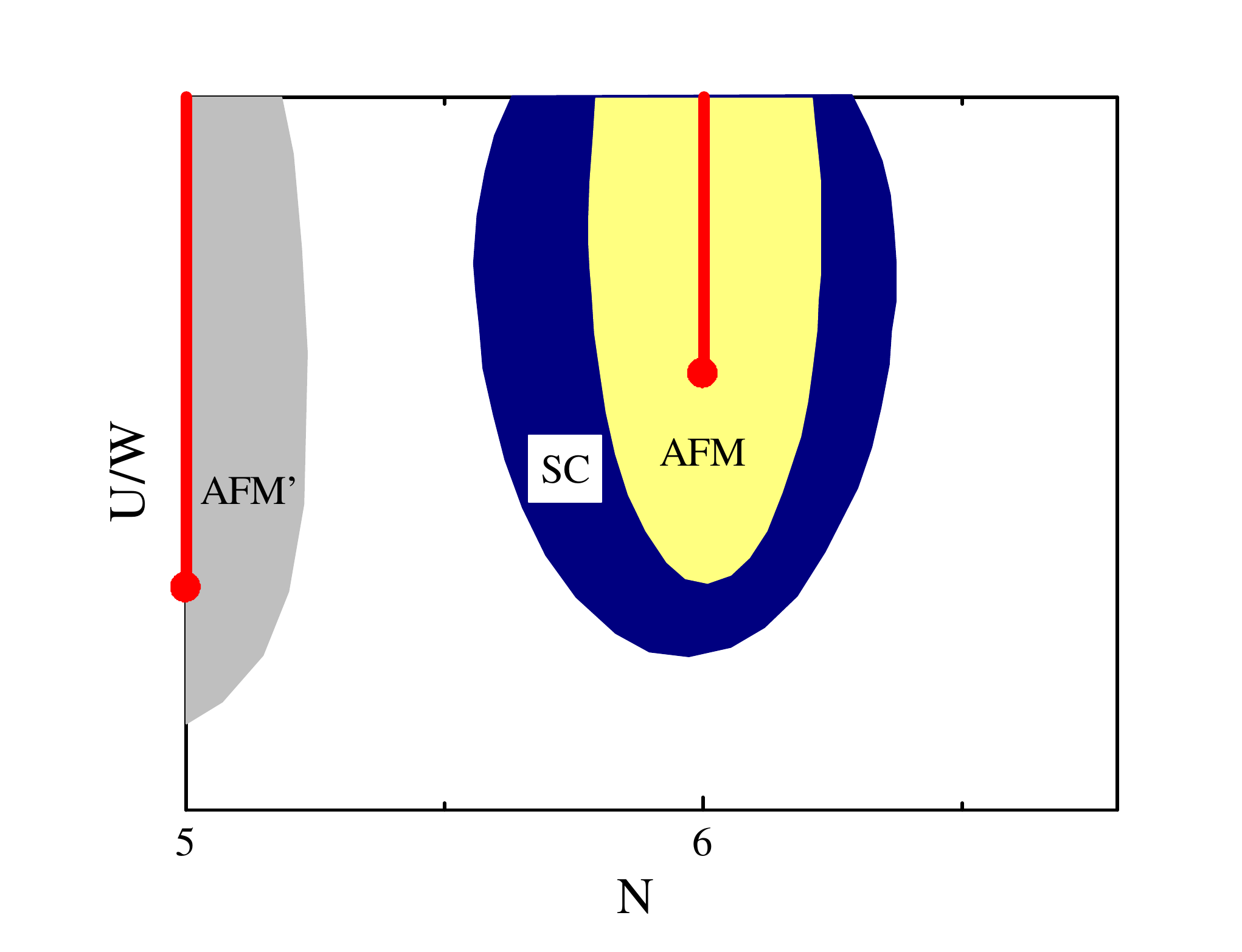}%
\vskip -1pc
\caption{\label{fig7} Sketch of the theoretical ground-state 
phase diagram of multiorbital models for iron-based superconductors beyond the $N=6$ electron occupancy.
The red solid lines at $N=5$ and $N=6$ refer to two Mott insulators at these respective fillings.
The red dots correspond to the points of the Mott transitions. They are respectively surrounded
by the antiferromagnetically ordered phases AFM and AFM$\prime$, which do not necessarily have the same
ordering wave vectors. The dark blue regime surrounding the AFM state corresponds to the superconducting
phase (SC), that is in proximity to both the antiferomagnetic order  and the Mott transition. }
\end{figure}

The dashed line in Fig.~\ref{fig8} illustrates a physical pathway that connects these two phases.
It corresponds to the combination of reduced strength of vacancy order and carrier doping.
ARPES measurements in an intermediate material, which contains
KFe$_{\rm 1.5}$Se$_{\rm 2}$ with
one vacancy per four iron sites \cite{Zhao12,MFang}
and a metallic phase with small carrier doping \cite{Yietal12,Chen11}, allows us to consider it 
as intermediate between the insulating K$_{0.8}$Fe$_{1.6}$Se$_2$ 
and the vacancy-free superconducting phase.
Along a similar line of consideration, it is possible that
 ${\rm K_{0.5}Fe_{1.75}Se_2}$, with
one vacancy per eight iron sites  \cite{Wen13},
can also be placed  along this physical trajectory.

\section{Going beyond the vicinity of $N=6$ electron occupancy}
%
%
\label{Sec:Discussion}

The systematic studies on the MIT of multiorbital models for parent iron-based superconductors discussed
 in previous sections confirm the existence of a Mott transition at commensurate filling $N=6$.
 With this Mott transition as the anchoring point, we can extend the discussion beyond the $N=6$ 
 electron occupancy
 (see Fig.~\ref{fig7}).
  In the strongly correlated metal regime, the incoherent electronic excitations give rise to
 quasi-localized magnetic moments; the corresponding spin physics
  can be described approximately by a $J_1$-$J_2$ model.
 These local moments exhibit a robust $(\pi,0)$ antiferromagnetic order near the Mott transition.
 Superconductivity develops near the boundary of the antiferromagnetic phase.

Recently, de'Medici \emph{et al.} has considered the orbital-dependent effects of correlated metals in proximity
to the Mott transition at $N=5$
~\cite{deMedici12}. Here we note that, due to the crystal field splittings, the high-spin configurations
for $N=5$ and $N=6$ are very similar except for the filling of the $3z^2-r^2$ orbital. Hence correlation
effects near the two Mott insulators should be similar. From the view of electron correlations, it could be more
convenient to take the $N=5$ Mott state as the parent compound for some heavily hole-doped iron-based
superconductors, such as KFe$_2$As$_2$. However, for the superconductivity in the electron-doped
and optimally (and under) hole-doped compounds, it is natural to consider that superconductivity
and the strong antiferomagnetic order near $N=6$ are anchored by the Mott transition at $N=6$.

\section{Summary and outlook}
\label{Sec:Summary}

We have summarized the theoretical investigations about the metal-to-insulator transition in multiorbital models
for the iron pnictides and alkaline iron selenides. In all these models, a Mott transition generally exists.
The nature of the Mott transition is strongly affected by the Hund's coupling. In particular, a strongly correlated
metallic state with suppressed quasiparticle spectral weights is stabilized within a wide range of model parameters
of the phase diagram. As a result of the proximity to the Mott transition, the system in the strongly correlated
metallic phase exhibits bad metal behavior. This highlights the importance of the correlation effects
in the iron pnictides. For alkaline iron selenides, the ordered vacancies drive the system from the bad metal
to a Mott insulator. As a unique feature of the interplay between the Coulomb/Hund's coupling and the
orbital non-degeneracy
in multiorbital systems, the electron correlations
lead to effects that are strongly orbital-dependent.
In alkaline iron selenides, this even yields a transition into an orbital-selective Mott phase.

The orbital-selective Mott behavior represents further evidence that electron correlations are important to the
microscopic electronic physics of the iron-based superconductors.
Equally important, the strong orbital selectivity highlights the multiorbital nature of the 
bad metal behavior of the iron-based superconductors. The orbital dependence of the correlation
effects connect well to both the experimental observations of the orbital-dependent effective mass
in the iron pnictides~\cite{Fujimori}, and the considerations of the orbital ordering ~\cite{Singh,Yi10}
in these materials. 
It will be important to explore how the orbital-dependent effect of the correlations
influence the nature of the superconducting state.
More generally, multi-orbital effects should also be important to other $d$ and $f$-electron systems at
the boundary of electron localization and itinerancy.

\section{Acknowledgment}

We would like to acknowledge Z. K. Liu, D. H. Lu, Z. X. Shen, L. L. Sun, and M. Yi for useful discussions.
This work has been in part supported by the NSF Grant No. DMR-1006985,
the Robert A. Welch Foundation Grant No. C-1411.
and  by the National Nuclear Security Administration of the U.S. DOE at LANL under Contract No.
DE-AC52-06NA25396 and the LANL LDRD Program.




\bibliographystyle{elsarticle-num}



\end{document}